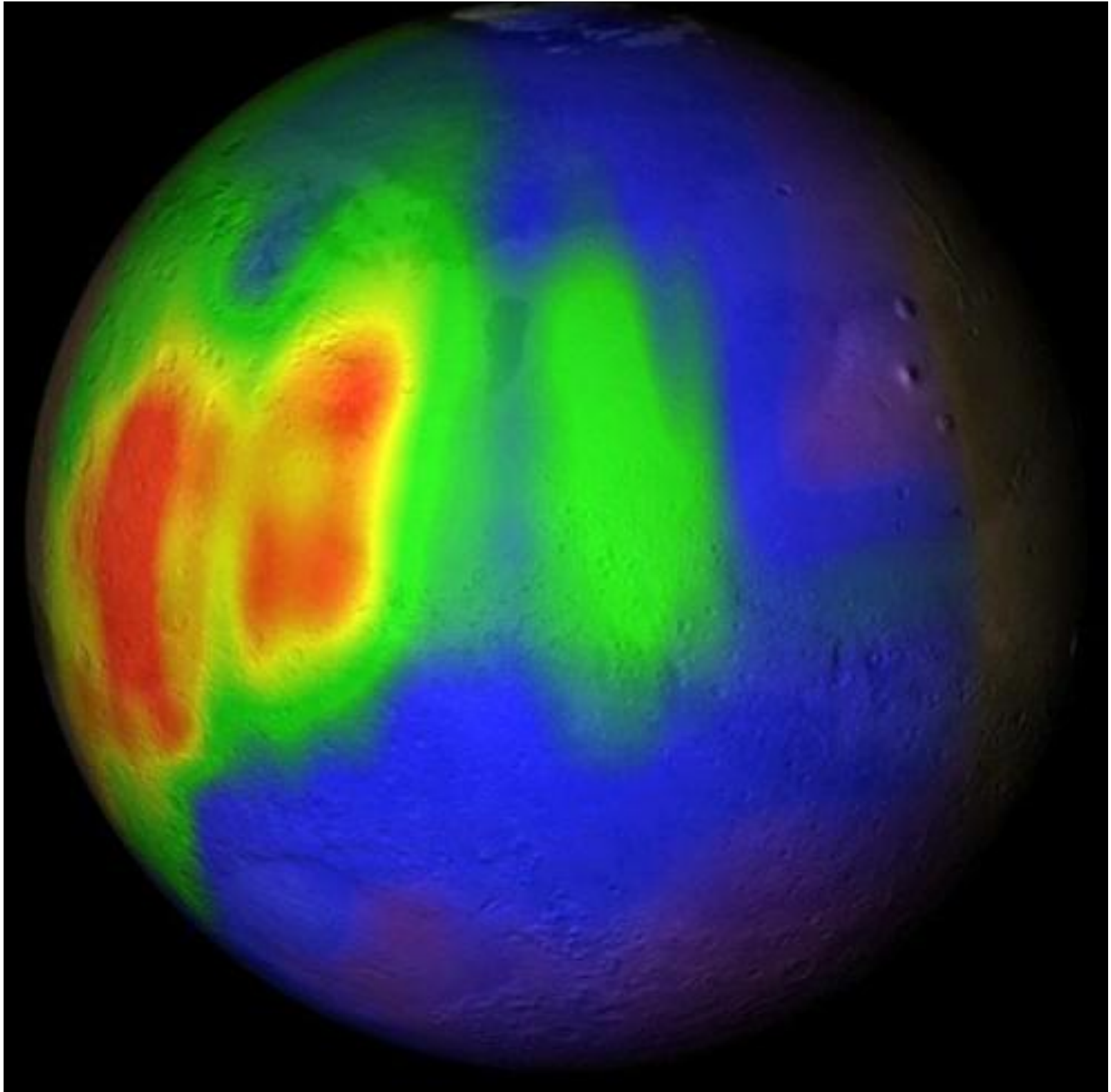

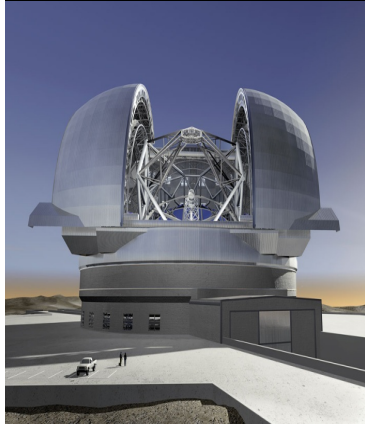
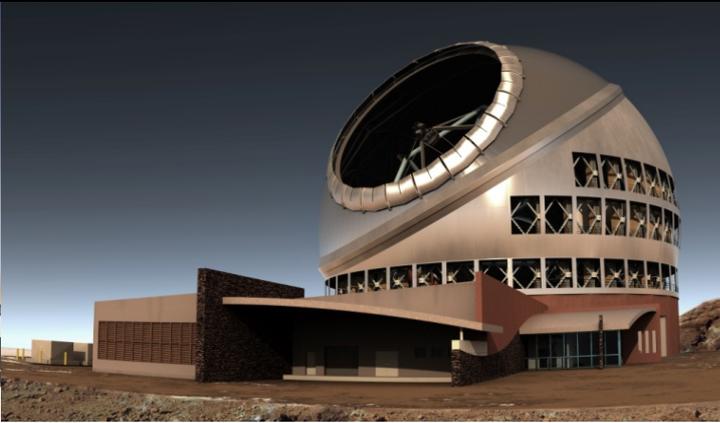
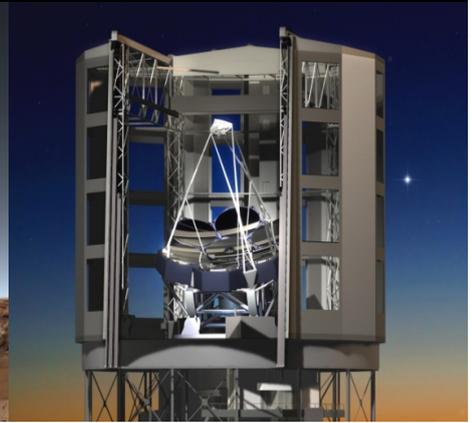

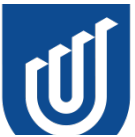
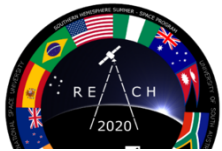
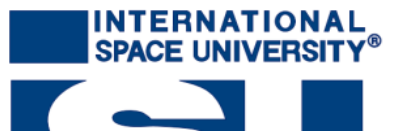

INTERNATIONAL
SPACE UNIVERSITY®

Potential for

Observing Methane on Mars

Using Earth-based

Extremely Large Telescopes

By

Robert Hunt

BAppSc (Curtin), GDipEd (Monash), GCertSc (Swinburne)

A Space Studies Project

as requirement for completion of a Graduate Certificate in Space Studies from UniSA

Cover Page Credits:

Mars Methane Map Courtesy NASA, E-ELT telescope courtesy ESO, TMT telescope courtesy TMT Observatory Corporation, GMT telescope courtesy GMT

TABLE OF CONTENTS

# ABSTRACT

The Red Planet has fascinated humans for millenia, especially for the last few centuries, and particularly during the Space Age. The nagging suspicion of extant Martian life is both fed by, and drives the many space missions to Mars, and recent detections of large, seasonal volumes of atmospheric methane (Mumma et al, 2003; Formisano et al, 2004; Chicarro, 2004; Krasnpolsky et al, 2004) have re-fueled the discussion. Methane's strongest vibrational frequency (around 3.3 μm) occurs in the lower half of astronomers' L Band in the near-infra red (NIR), and is readily detectable in the Martian atmosphere from ground-based spectrographs at high, dry locations such as Hawaii and Chile. However, resolution of specific spectral absorption lines which categorically identify methane are disputed in the literature, as are their origins (Encrenaz, 2008; Zahnle, 2011). With the proposed construction of Extremely Large Telescopes (ELTs) operating in the optical/NIR, the question became: could these ELTs supplement, or even replace space-based instruments trained on Martian methane? A review of immediate-past, present, and future NIR spectrometers on-Earth, in-air, in Earth orbit, in solar orbit, in L2 orbit, in Mars orbit, and on Mars, revealed a wide range of capabilities and limitations. Spatial, spectral, radiometric, and temporal resolutions were all considered and found to be complex, inter-related and highly instrument-specific. The Giant Magellan Telescope (GMT), the Thirty Meter Telescope (TMT), and the European Extremely Large Telescope (E-ELT) will each have at least one L Band NIR spectrometer supported by state-of-the-art adaptive optics, and capable of extreme spatial, spectral and radiometric resolution. Replicating observations over time will provide a critical constraint to theoretical considerations about the biotic or abiotic origins of any detected methane, and it is recommended that existing datasets be mined, science cases for the ELTs include Martian methane, and collaboration between science teams be enhanced.

# ACKNOWLEDGEMENTS

This review would not have been possible without the guidance and support of the co-ordinators of the International Space University's Southern Hemisphere Summer Space Program held at the University of South Australia. They were Research Associate Professor, Dr Scott Madry from the University of North Carolina, and Associate Professor, Dr David Bruce from the School of Natural and Built Environments at the University of South Australia.

Particular thanks go to my immediate supervisor, and Mars Society colleague, Dr Jonathan Clarke, from Geoscience Australia and the Australian Centre for Astrobiology at the University of New South Wales, Australia.

Thanks also go to specialists who were kind enough to provide answers to many questions along the way:

- Prof., Dr Chris Tinney - School of Physics, University of NSW, Australia
- Assoc. Prof., Dr Jeremy Bailey - School of Physics, University of NSW, Australia
- Prof., Dr Gary Da Costa - Research School of Astronomy and Astrophysics at The Australian National University, Canberra, Australia
- Dr Chis Flynn - Centre for Astrophysics and Supercomputing at Swinburne University, Melbourne, Australia
- Dr Hans-Ulrich Kaufl, Principle Investigator, CRIRES, European Southern Observatory, Chile
- Assoc. Prof. of Astronomy, Dr Bernard Brandl, Leiden University, The Netherlands
- Dr Stuart Ryder, Australian Gemini Scientist, Australian Astronomical Observatory
- Joshua Nelson - Chief Engineer at Mars Desert Research Station, Utah, USA
- Michael Kueppers - Scientist in the ESA Rosetta Science Operations Team and OSIRIS Co-Investigator, ESA Directorate of Science and Robotic Exploration
- Michael Smith, NASA Goddard Space Flight Center

# LIST OF FIGURES

# LIST OF ACRONYMS

| | |
|---|---|
| **A** | |
| AO | Adaptive Optics |
| **B** | |
| BASS | Broadband Array Spectrograph System |
| BOLD | Biological Oxidant and Life Detection |
| **C** | |
| CCD | Charge Couple Device |
| CH$_4$ | Methane |
| CRIRES | Cryogenic High-Resolution InfraRed Echelle Spectrograph |
| CRISM | Compact Reconnaissance Imaging Spectrometer for Mars |
| CSHELL | Cryogenic Echelle Spectrograph |
| **D** | |
| DN | Digital Number |
| **E** | |
| E-ELT | European Extremely Large Telescope |
| ELT | Extremely Large Telescopes |
| ESA | European Space Agency |
| **G** | |
| GMT | Giant Magellan Telescope |
| GMTNIRS | Giant Magellan Telescope Near Infra Red Spectrometer |
| GOWON | Gone With The Wind On Mars |
| GSFC | Goddard Space Flight Center |
| **F** | |
| FISTA | Flying Infrared Signatures Technology Aircraft |
| FLITECAM | First Light Infrared Test Experiment Camera |
| FTS | Fourier Transform Spectrometer |
| FTIS | Fourier Transform Infrared Spectroscopy |
| FWHM | Full Width Half Maximum |
| FWC | Full Well Capacity |

**G**

| | |
|---|---|
| GMTNIRS | Giant Magellan Telescope Near InfraRed Spectrometer |
| GNIRS | Gemini Near Infra Red Spectroscope |
| GOWON | Gone With The Wind On Mars |

**H**

| | |
|---|---|
| HITRAN | High Resolution Transmission Molecular Absorption |
| HST | Hubble Space Telescope |

**I**

| | |
|---|---|
| IPAC | Infrared Processing and Analysis Centre |
| IRAC | InfraRed Array Camera |
| IRC/NIR | Infra Red Camera/Near Infra Red |
| IRIS | Infra Red Imager and Spectrometer |
| IRTF | Infra Red Telescope Facility |
| ISAAC | Infrared Spectrometer And Array Camera |
| iSHELL | Immersion Grating Echelle Spectrograph |
| ISO | Infrared Space Observatory |
| ISOCAM | Infrared Space Observatory Camera |
| ISOPHOT | Infrared Space Observatory Photo-polarimeter |
| ISU | International Space University |
| IUPAC | International Union of Pure and Applied Chemistry |

**J**

| | |
|---|---|
| JWST | James Webb Space Telescope |

**K**

| | |
|---|---|
| KPNO | Kitt Peak National Observatory |

**L**

| | |
|---|---|
| LCROSS | Lunar Crater Observation and Sensing Satellite |

**M**

| | |
|---|---|
| MAVEN | Mars Atmosphere and Volatile Evolution Mission |
| METIS | Mid-infrared E-ELT Imager and Spectrograph |
| MIISE | Mid-IR Imaging Spectrograph |
| MIRIS | Midwave Infrared Imaging Spectrograph |
| MGS | Mars Global Surveyor |
| MRO | Mars Reconnaissance Orbiter |
| MSL | Mars Science Laboratory |
| **N** | |
| NASA | United States National Aeronautics & Space Administration |
| NGIMS | Neutral Gas and Ion Mass Spectrometer |
| NICMOS | Near Infrared Camera and Multi-Object Spectrometer |
| NIR | Near Infra Red |
| NIRC | Near Infra Red Camera |
| NIRC2 | Near Infra Red Camera 2 |
| NIRCAM | Near Infrared Camera |
| NIRES | Near Infra Red Echelle Spectrometer |
| NIRISS | Near-InfraRed Imager and Slitless Spectrograph |
| NIRSPEC | Near Infrared Spectrometer (Keck) |
| NIRSpec | Near InfraRed Spectrograph (JWST) |
| NSF | United States National Science Foundation |
| NSFCAM2 | National Science Foundation Camera 2 |
| **O** | |
| OMEGA | Visible and Infrared Mineralogical Mapping Spectrometer |
| **P** | |
| PAH | Polycyclic Aromatic Hydrocarbon |
| PFI | Planet Formation Instrument |
| PFS | Planetary Fourier Spectrometer |
| **S** | |
| SAIRS | Synthetic Aperture Infra Red Spectrograph |
| SAO | Smithsonian Astrophysical Observatory |
| SEDS | Students for the Exploration and Development of Space |
| SHS-SP | Southern Hemisphere Summer Space Program |
| SNR | Signal to Noise Ratio |
| SOFIA | Stratospheric Observatory for Infrared Astronomy |
| SPeX | Spectrograph and Imager |
| SWS | Short Wave Spectrometer |
| **T** | |
| TES | Thermal Emission Spectrometer |
| TLS | Tunable Laser Spectrometer |
| TMT | Thirty Meter Telescope |
| **U** | |
| UniSA | University of South Australia |
| USAF | United States Air Force |
| **V** | |
| VIRTIS | Visible and Infra Red Thermal Imaging Spectrometer |
| VLTA | Very Large Telescope Array |
| VPH | Volume Phase Holographic |
| **W** | |
| WIRC | Wide Field AO Imager |
| WISE | Wide Field Infrared Survey Explorer |

# 1.    INTRODUCTION

## 1.1    Scope

The International Space University's (ISU) 2012 Southern Hemisphere Summer Space Program (SHS-SP12) was conducted in partnership with the University of South Australia (UniSA). This paper constituted the remaining requirements to complete a Graduate Certificate of Space Studies from UniSA, and consistent with the SHS-SP12 remote sensing theme, reviewed near infra red astronomical spectroscopy, and the potential for ground-based, extremely large telescopes to do meaningful atmospheric science of Mars.

## 1.2    Background

The birth of remote sensing is commonly attributed to Gaspard-Félix Tournachon's 1858 aerial photographs of Paris from a hot air balloon. But if remote sensing is the activity of using light-sensitive tools to record details of distant objects, then astronomers have been doing it since Galileo.

Currently in design phase are three optical/infra-red, extremely large, Earth-based telescopes in the order of 30 m – 40 m diameter. The Thirty Meter Telescope (TMT) will be built on Mt Maunakea in Hawaii (TMTweb, 2012), the Giant Magellan Telescope (GMT) and the European Extremely Large Telescope (E-ELT) are planned for the Atacama, Chile (E-ELTweb, 2012). These telescopes will incorporate the latest in design and technology for mirrors, adaptive optics, electronics and structural engineering.

Given recent budget cuts suffered by NASA (United States National Aeronautics & Space Administration) and ESA (European Space Agency), it's relevant to consider the future use of these resources to study Mars. Extremely Large Telescopes (ELTs) are an obvious option.

Modern spectrometers on ground-based telescopes can detect many chemical species on Mars (Villanueva et al, 2011, Radeva et al (a), 2010, Encrenaz, 2004, Novac et al, 2011, Redeva et al (b), 2010)  and can subtract the effects of Earth's intervening atmosphere ('telluric' spectral lines) to draw accurate conclusions about the target. The proposed ELTs will have spectrometers which will extend astronomers' capabilities to define, amongst many other things, the chemical composition of atmospheres and surfaces of solar system bodies, including Mars.

Remote sensing technologies consider four types of target resolution. Temporal (over time), radiometric (bit data), spatial (over distance), and spectral resolution (across wavelengths) all reveal important information about a target, usually with trade-offs between each. With respect to methane detection on Mars, the strongest absorption line is around the 3.3 μm ($9.1 \times 10^{13}$ Hz or 3030 cm$^{-1}$) in the Near Infra-red (NIR), 'L' Band, and all three of the proposed new telescopes have at least one spectrometer which can operate within this band. Consideration was also given to detections of other methane lines at 1.7 μm, 2.3 μm, 6.3 μm and 7.7 μm.

1.3    Basics of Astronomical Spectroscopy

The basic components of an astronomical spectrometer (Figure 1) are the telescope, used to collect incoming light, the slit used to exclude non-target light, the collimator to introduce parallel light rays into the dispersing element, which separates light into a spectrum, and a camera to focus the output onto a detector (Kilkennyweb, 2012). There are various technologies for each of these parts, designed to match the telescope and science drivers.

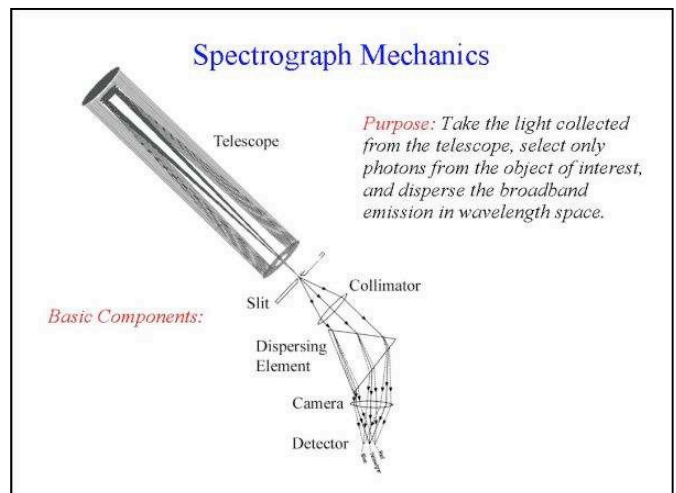

Figure 1 Basic Components of a Spectrograph      Credit: Kilkenny

Telescopes are typically a type of Cassegrain design using hyperbolic mirrors, light from the source reflecting off the primary, onto a secondary mirror and subsequently back through the center of the primary, coming to a focal plane. Light capture is important, so diameter (aperture) is an observational constraint.

At the focal plane may be placed a variety of obscuring masks (often a slit) to block non-source light, the length and width of the resultant aperture being an important parameter of the system.

Diverging light rays exiting the slit are straightened into a parallel beam by a collimator designed to optimally capture the incident light, else the system loses light or resolution. The now parallel light rays are directed onto a dispersing apparatus.

There are a number of technologies for spreading light into its spectrum. The original and perhaps most common method uses a glass prism, though diffraction gratings are now widely used when a specific wavelength is being observed. A particular type of diffraction grating, known as an Echelle, has fewer grooves per millimetre, and uses a technique called 'cross-dispersion' to achieve high resolution. Slitless spectroscopes use a 'grism', which is a combination of a prism and a diffraction grating, that allows the collimated target wavelength to pass through it (STSIweb, 2012).

The capability to separate wavelengths of light is resolving power $R = \frac{\lambda}{\Delta\lambda}$ but the spectral resolution of the overall system is also influenced by the detector slits, image aberrations, and magnification (Newportweb, 2005).

Long-slit spectroscopes improve signal-to-noise ratio by excluding unwanted source light. By 'nodding' the telescope up and down within the slit, they have the advantage of being able to compare the spectrum of a small light source with the sky either side of it, thereby subtracting the sky from the final spectrograph (DurhamUweb, 1998). It is also used to obtain spatial and spectral data at the same time, by recognising Doppler shifts in the resultant spectrograph (Cornellweb, 2012).

Multi-object spectroscopy uses many optical fibres at the focal plane to simultaneously detect multiple sources. It is therefore a slitless spectroscopy, each optical fibre directing its input to a detector (NICMOSweb, 2012).

Integral Field (also known as 2- or 3-dimensional) spectroscopy uses a complex combination of lenslets, optical fibres and image slice processing to produce a spectrum from many parts of the target at the same time (DurhamUweb, 1998).

Raman spectrometers excite a target's atoms with a laser beam and measure the change in the beam's energy level. Tuneable laser technology is a refinement of Raman technology, using a precise wavelength semiconductor laser to initially excite a specific electron transition in a target chemical species, then test the target's response at wavelengths slightly higher and lower than its fundamental vibrational frequency. Modern tuneable lasers need not operate at cryogenic temperatures (TLSDweb, 2012).

A spectrometer's camera usually re-focuses the output of the dispersing element onto a detector, in most cases a Charge Coupled Device (CCD). As well as camera optics, the physical and electronic specifications of the CCD strongly determine the sensitivity and resolution of the final spectrograph. In-situ calibration of the CCD pixels can be achieved by comparing a laboratory standard spectrum of the target species immediately before and after a source reading. Techniques such as 'nod and shuffle' whereby CCD array charges are shuffled in synchronisation with the telescope's nodding, can improve signal to noise ratio significantly (Glazebrook et al, 1998).

1.4     Martian Atmospheric Methane

Mariner 7 had an Infra Red Spectrometer which apparently detected Martian atmospheric methane at 3.3 μm, in 1969. Following an public announcement, the discovery was withdrawn because it was clear that the detection was likely $CO_2$ ice (Sullivan, 1969). Since a controversial 2003 detection (Mumma et al, 2003) of high levels of Martian atmospheric methane by a team from NASA's Goddard Space Flight Centre (Figure 2), the issue was re-ignited, and there have been hundreds of peer-reviewed journal papers written on the subject. Various scientists derived clear, though contradictory conclusions regarding a cause, but in a recent summary of the last decade, Zahnle (2011) finds that better, more accurate science can be, and should be done, to resolve this question, both from ground-based facilities and from Mars missions.

Oliva and Origlia (2008) stated that high resolution, near infrared spectroscopy was a new area of interest and that contemporary instruments had surprisingly poor spectral resolution, recommending a newly designed instrument for the European Extremely Large Telescope (E-ELT). This highlighted the scientific novelty of the NIR, and validated the premise of the current work.

Incident infrared radiation is absorbed by $H_2O$ and $CO_2$ in the Earth's atmosphere, though some wavelength windows do exist. The atmosphere itself emits infrared radiation, compounding the problem for ground-based telescopes, but the lower part of astronomers' $3.0 - 4.0$ μm 'L' Band has fairly good seeing conditions at high altitudes (IPAC, 2012a). Telluric water vapour and methane, can be addressed "… by ratioing the spectrum of the science target by that of a calibration star…" (Gemini, 2012) and 'guide stars' can be artificially created with lasers. Telluric methane is commonly subtracted from spectrographs using the HITRAN (high-resolution transmission molecular absorption) database which quantum characterises them during observations (HITRAN, 2012).

Methane has four spectral bands at around 7.3 μm, 6.3 μm, 3.3 μm and 3.2 μm (ULiv, 2012). However, Formisano et al (2004) used the Planetary Fourier Spectrometer (PFS) at around 7.7 μm, 3.3 μm, 2.3 μm and 1.7 μm. Regardless, the resonant $v_3$ band at around 3.3 μm has the strongest signal to noise ratio, is most easily distinguished from other effects, and is the focus of much of the literature on Martian methane.

Temporal resolution refers to the rate of return to a target site. From Earth, depending on relative orbital positions, Mars varies from being 56 to 400 million Km away (GSFC, 2012), and with a diameter of 6760 Km, presents an angular size 3.5" – 25". So Mars is best viewed when close to Earth, that is, at opposition. This occurs approximately every 26 months, and upcoming oppositions present large diameter Martian discs: 15.2" in 2014, 18.6" in 2016, 24.3" in 2018, 22.6" in 2020, and 17.2" in 2022 (SEDSweb, 2012). Since Mars' axial rotation is only slightly different to Earths, terrestrial observations of a particular location on Mars can occur over about a sixteen day period, after which it takes another three weeks for the same location to reappear on the disc (MarsProfilerweb, 2012).

Mars Orbiters, while repeating their orbit about every one hundred minutes, do not return over the same site for many months, and Martian rovers can only sample their immediate environment over their mission life time. Spacecraft in other orbits (solar, L2 etc), have different periodic access to a Martian apparition, also depending on orbital specifics.

Pixels in Charge Couple Devices (CCDs) collect electrons up to their Full Well Capacity (FWC) - as many as 500,000 each, and radiometric resolution relates to the FWC of the CCD and the digitisation process, giving a Digital Number (DN). The FWC/DN ratio is the CCD's gain and is related to the CCD array size (Specinst, 2012; RITweb, 2012).

Instrument sensitivity and radiometric resolution is "…a very complex field…" (Brandl, 2012), detector responsivity and atmospheric background determining overall radiometric resolution. Astronomical instruments use 16-bit pixel resolution which is adjusted for a particular FWC for the wavelength being detected (Ryder, 2012). This was corroborated by Seppo (2012) who regarded DN as the basic radiometric resolution, as opposed to Smith (2012) who uses FWC and Signal to Noise Ratio (SNR). Internal and external noise are the greater contributors to limiting radiometric resolution, and these are ameliorated by taking multiple, short detections, as well as 'dark frame' (closed aperture) detections, and averaging the result. Environmental thermal background 'noise' is addressed by locating telescopes at high, dry altitudes and using appropriate structural materials and coatings (Ryder, 2012).

"…the history of the search for methane on Mars is long and storied." (Mumma, 2004). The infant technology of spectroscopy was applied to Mars from the mid 19[th] century (Campbell, 1894) and later, Kuiper et al (1947) thought that a newly designed spectrograph could help solve the nature of "…the green spots on Mars…". 'Sinton Bands' were controversially thought to be caused by methane as recently as 1957 (Sinton, 1957).

Detection of trace amounts of methane in Mars' atmosphere by Mariner 9 (Maguire, 1977; Hartmann & Raper, 1974) were corroborated telescopically in 1997 (Krasnopolsky, 1997; Krasnopolsky et al, 1997), and subsequently Summers et al (2002) suggested volcanic outgassing, Wong et al (2003) favoured biogenesis, and Caldwell et al (2003) showed how the Hubble Space Telescope (HST) could make useful observations. But it was Mumma et al (2003, 2004) who detected significant amounts with the CSHELL (Cryogenic Echelle Spectrograph) and Phoenix instruments on telescopes at Mt Maunakea.

Mars Express confirmed this detection with its Planetary Fourier Spectrometer (Formisano et al, 2004; Chicarro, 2004), and Krasnopolsky et al (2004) made a similar confirmation using the Fourier Transform Spectrometer at the Canada-France-Hawaii Telescope, suggesting a biological cause. Steigerwald (2009) explained that thousands of tonnes of methane in the Martian atmosphere points to a recent geological or biological process.

Discussions have ranged from production of methane by methanogens in a laboratory (Kral et al, 2003), to cometary impact, hydrothermal processes, or extant micro-organisms (Atreya & Wong, 2004). Lyons et al (2005) promulgated subsurface magma interactions, Krasnopolsky (2006) was unequivocal about flaws in the abiotic case, though a year later (Krasnopolsky, 2007) could not detect any methane using the same instrument as Mumma. Martin-Torres and Mlynczak (2005) suggested solar heating was a complicating factor. Mumma et al (2009) published data showing tens of thousands of tonnes of methane, Encrenaz (2008) casting doubt over the detections suggesting that "…further dedicated observations are … needed to firm up the detection …". Spacecraft detectors have even been designed which promise to successfully distinguish between biogenic and geologic methane (Wilson et al, 2007).

It was not until three years ago that high dispersion infrared spectrometers with spatial resolutions better than 500 Km, were aimed at Mars (Mumma et al, 2009), and because of Mars' thin atmosphere, spectral line broadening due to pressure is minimal, so high spectral resolution is required to separate atmospheric species (Encrenaz, 2008).

Mumma's group maintain that their ground-based readings (up to 60 ppb) using Doppler-shifted spectra show an unexpectedly short 'seasonal' and 'latitudinal' variation of methane on Mars. The variability is most interesting, and Mars Express collected corroborating data (Formisano et al, 2004). Methane on Mars is well understood to need several hundred years to dissipate chemically, whereas contemporary observations indicate transience on a scale of weeks. The problem is that there is no known Martian atmospheric chemistry which allows such volumes of methane to be oxidised or condensed away, without an accompanying release of chemical byproducts into the atmosphere (Zahnle, 2010). Fonti and Marzo (2010) agree that the lines seen by Mumma's group are likely from a telluric isotopologue ($^{13}CH_4$) of methane (Figure 2).

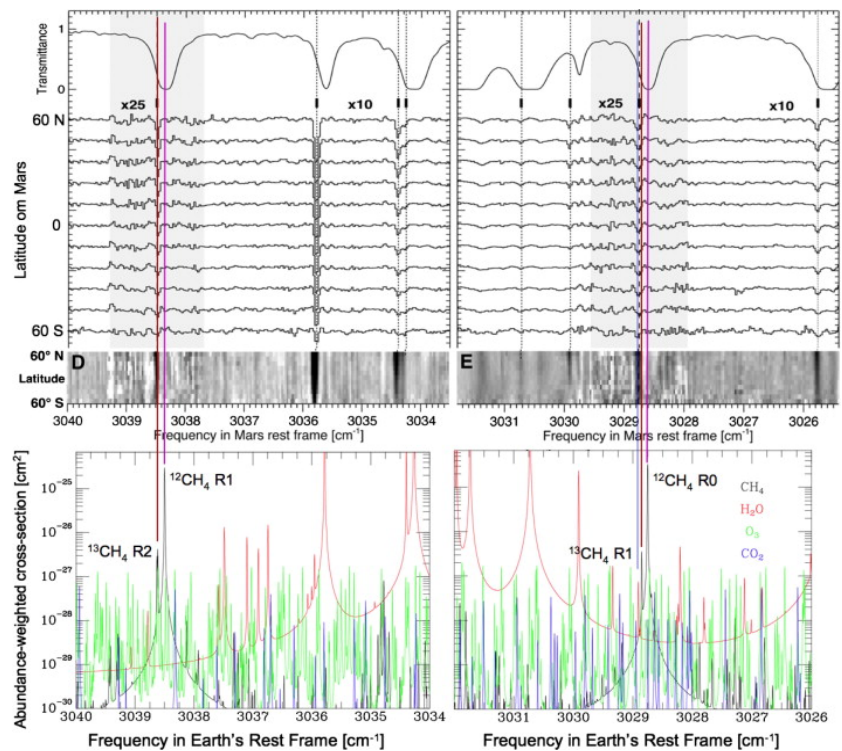

Figure 2 Apparent methane detection by Mumma (upper chart) corresponds very closely to blue-shifted telluric $^{13}CH_4$ lines (lower chart)    Credit: Zahnle

The biogenic argument is also criticised by Zahnle (2010) who reminded us that extant Martian methanogens using CO from the atmosphere (a well understood terrestrial ecology) would re-supply atmospheric methane while depleting the CO. But the Martian atmosphere contains plenty of CO.

Zahnle (2010) also warns against the specifics of using HITRAN as an atmospheric subtraction tool. Given a terrestrial atmospheric concentration of methane around 1,800 ppb, and a Martian concentration around, say, 20 ppb, then using Doppler shifted spectra can distinguish the *peaks* of the respective lines, but can't remove the Martian lines from the wings of the telluric lines which are much broader. This is further muddied by the aforementioned telluric $^{13}CH_4$ lines which are much stronger than the Martian $^{12}CH_4$ lines and close to them in the spectrum.

Zahnle (2010) recommends doing much better science both from Earth and from space. The use of bigger terrestrial telescopes with more powerful spectrographs, the use of SOFIA (Stratospheric Observatory for Infrared Astronomy) to do observations above the atmosphere, and the potential of the current Mars Science Laboratory's onboard laser spectrometer, are all recommended.

## 1.5 The Three ELTs

A Giant Magellan Telescope (GMT) science case (GMTweb, 2006) included studying solar system objects with 'more sensitive spectroscopy'. By 2008, the GMTNIRS (Giant Magellan Telescope Near Infra Red Spectrometer) was proposed, looking in the $3 - 5$ μm range with spectral resolution up to 150,000 and spatial resolution 30 milliarcseconds (mas) (GMTweb, 2008). The GMT will have capacity to observe Mars in the $1 - 5$ μm range with high spectral resolution and small field of view with respect to the disk of Mars (Tinney, 2012; Da Costa, 2012). Ground has recently been broken on construction (GMTBlastweb, 2012).

A Thirty Meter Telescope (TMT) science case (TMTweb, 2007) suggested their Infra Red Imager and Spectrometer (IRIS) could examine the chemical makeup of Jovian moon atmospheres to study the evolution of conditions suitable for life on Earth: a science driver for Martian exploration. Similar to the GMTNIRS, the TMT will also have a Near-InfraRed Echelle Spectrometer (NIRES) (TMTweb, 2010). Although time on the telescope is 'over-subscribed', periodic, short observations of 15 minutes was considered a possibility (Ádámkovics, 2012).

The European Extremely Large Telescope (E-ELT) science document (Hook, 2007) explicitly mentions solar system astronomy, espousing the virtues of a telescope with spatial resolution similar to Earth's weather satellites, proposing great benefits from temporal observations of the planets, complementing the work of spacecraft missions. It was subsequently argued that the E-ELT would make important contributions to near-by planetary astronomy, by virtue of its spatial, spectral and temporal resolution (E-ELTweb, 2009). The Mid-infrared E-ELT Imager and Spectrograph (METIS) will have spectral resolution up to 100,000 in the $3 - 5.3$ μm band. Budgetary approval has been granted for initial groundworks.

## 2.    AIMS AND OBJECTIVES

The aim of this project was to explore the value of Earth-based ELTs as a tool for observing Martian atmospheric methane.

The objectives were to:

2.3.1    document the relevent capabilities and limitations of the three proposed ELTs,

2.3.2    summarize the capabilities of instruments at other locations, and

2.3.3    recommend future Mars science drivers for the ELTs

## 3.    METHODOLOGY

This project used internet research to review operational specifications of over forty past, present, and proposed NIR spectrometers capable of detecting 3.3 μm. These details were synthesised into a discussion with respect to the suitability of the proposed ELTs for informing the science of Martian atmospheric methane.

# 4.    RESULTS AND DISCUSSION

This investigation identified dozens of NIR spectrometers capable of detecting methane's 3.3 µm molecular vibration. Appendix 6.2 sorts them according to location: on Earth, in flight, Earth orbit, solar orbit, orbiting L2, Mars orbit, and on Mars. Instrument titles were hyperlinked to a website containing their specifications. Not an exhaustive list, it contains spacecraft/instruments which have ceased operations, and some which are yet to be built/launched, including the three ELTs. Sections 4.1 to 4.7 provides an overview of instruments at those locations. Section 4.8 presents results for the concepts of instrument sensitivity, as well as a discussion of each of the four types of resolution: spatial, spectral, radiometric and temporal.

## 4.1    NIR Spectrometers on Earth

Not counting the proposed ELTs, at least thirteen instruments on four terrestrial platforms were described. The most spectrally sensitive spectrometers currently in use were CRIRES (Cryogenic High-Resolution InfraRed Echelle Spectrograph) at VLTA (Very Large Telescope Array) with R = 100,000, and iSHELL (Immersion Grating Echelle Spectrograph) at IRTF (Infra Red Telescope Facility) with R = 67,000. CRIRES had a better spatial resolution than iSHELL, and also detects methane at its lower vibrations around 1.7 µm and 2.3 µm.

Keck Observatory's NIRC (Near Infra Red Camera) has been superceded by NIRC2 (Near Infra Red Camera 2) which was claimed to have the highest spectral resolution of any terrestrial instrument (though a figure was not found), a specific 3.3 µm filter, and declared science goals including studying the solar system. With 0.01" per pixel on a 1024 x 1024 array, this promises to be a very useful instrument.

The Gemini Observatory at Mt Maunakea has GNIRS (Gemini Near Infra Red Spectroscope), with low spectral resolution, but very selective wavelegth filters around 3.3 µm. And Phoenix, an upgraded Gemini instrument which will deliver very high spectral resolution around 70,000, will be used in Arizona.

Also at Mt Maunakea, IRTF's NSFCAM2 (National Science Foundation Camera 2) has a large CCD (2048 x 2048 like iSHELL), though poorer spectral resolution, but is recommended over SPeX (Spectrograph and Imager) for L Band viewing because of its good spatial resolution. BASS (Broadband Array Spectrograph System) and MIRIS (Midwave Infrared Imaging Spectrograph) are portable instruments which could be used aboard SOFIA to take broad spectrum images at low spectral resolution.

For the ELTs, there are a total of seven proposed instruments. The GMT will have GMTNIRS, an extremely high spectral resolution instrument (R = 120,000), and a low resolution instrument (MIISE, Mid-IR Imaging Spectrograph) which will cover a large area of the sky.

The TMT will have four instruments of interest, all using very large 4096 x 4096 pixel CCDs. With spectral resolution up to 100,000 and pixel capture of just 0.0035", NIRES will be an extremely sensitive tool, whereas WIRC (Wide Field AO Imager) and PFI (Planet Formation Instrument) will be low spectral resolution, wide field detectors. IRIS will use multiplexed optical fibre and may be able to observe methane's lower wavelength vibrations.

The E-ELT has proposed one instrument of interest, METIS, with very high spectral resolution of 100,000, which is still in early design phase.

4.2     NIR Spectrometers in Flight

FLITECAM (First Light Infrared Test Experiment Camera) aboard SOFIA has already done interesting science, and a proposal to study exoplanet atmospheres using its tailor-made 3.3 μm filter falling onto a large 2048 x 2048 pixel CCD, as well as the 2.7 m aperture, make it a very powerful tool for above-atmosphere NIR science.

4.3     NIR Spectrometers in Earth Orbit

Of the eight NIR instruments found in Earth orbit, four have ceased operations (the Infrared Space Observatory Camera, the Infrared Space Observatory Photo-polarimeter, and the Short Wave Spectrometer on the Infrared Space Observatory; and the Infra Red Camera/Near Infra Red on Akari), two are dedicated all-sky survey missions (InfraRed Array Camera on the Spitzer Space Telescope, and the Wide Field Infrared Survey Explorer (WISE)), one, NICMOS (Near Infrared Camera and Multi-Object Spectrometer) on HST (Hubble Space Telescope), can only detect the weaker 1.7 µm and 2.3 µm lines for methane, and one is a 'blue sky' proposal (Arkyd). All have small CCDs and low spatial resolutions, but IRAC and WISE in particular can detect atmospheric methane at great distances.

The datasets for all these instruments may contain valuable information either intrinsically, or in support of other observations, and could be mined in the future. Also, NICMOS, IRAC, WISE, and Akryd may be used for repeat observations of Mars, to corroborate detections by other instruments. They have the distinct advantage of being above Earth's atmosphere.

## 4.4    NIR Spectrometers in Solar Orbit

VIRTIS (Visible and Infra Red Thermal Imaging Spectrometer) is onboard ESA's Rosetta, is the only NIR instrument in solar orbit, and will flyby Mars around 2015. It is an instrument designed to detect gases around solar system bodies, so if still functioning, this presents an opportunity to take high resolution methane spectra of Mars at 3.3 µm.

## 4.5    NIR Spectrometers Orbiting L2

With high spatial resolution, moderate spectral resolution, a prime viewing position, and potential for repeated observations, the three NIR instruments on the James Webb Space Telescope (the Near Infrared Camera, the Near InfraRed Spectrograph and the Near-InfraRed Imager and Slitless Spectrograph) could provide exciting observations of Martian atmospheric gases. With Mars now gradually returning to a close approach to Earth in opposition, in 2018, JWST would be 1.5 million Km closer to its target than ground-based, or Earth-orbiting telescopes, and without telluric contamination.

## 4.6    NIR Spectrometers Orbiting Mars

There are currently three instruments orbiting Mars on two spacecraft. OMEGA (Visible and Infrared Mineralogical Mapping Spectrometer) on Mars Express is targeting the Martian surface, but does have good spatial resolution (300 m) across a broad bandwidth. PFS on Mars

Express is specifically designed to characterise the Martian atmosphere covering several methane lines, and with the capability of collecting millions of spectra over time, is a powerful tool.

The third instrument in orbit is CRISM (Compact Reconnaissance Imaging Spectrometer for Mars) onboard MRO (Mars Reconnaissance Orbiter), and it has similar specifications and potential to PFS.

Although the MGS (Mars Global Surveyor) mission has ended, millions of 7.8 µm spectra collected over five years from its TES (Thermal Emission Spectrometer) instrument is enabling scientists (Fonti & Marzo, 2010) to map seasonal changes in methane at this wavelength.

Scheduled to launch in the next year or two, NASA's MAVEN (Mars Atmosphere and Volatile Evolution Mission) will use the as-yet-unbuilt NGIMS (Neutral Gas and Ion Mass Spectrometer) to definitively answer the Martian methane question.

4.7     NIR Spectrometers on Mars

Notwithstanding 'blue sky' ideas such as GOWON (Gone With The Wind On Mars) and BOLD (Biological Oxidant and Life Detection), there is only one mission designed to measure methane on Mars in-situ. The Mars Science Laboratory's TLS (Tunable Laser Spectrometer) will provide extremely high precision, on-ground measurements in the order of parts per trillion. MSL is due to land in August, 2012.

4.8     Resolution and Sensitivity

Astronomical, NIR spectroscopy at 3.3 µm has been, is being, and will be, conducted from a multitude of platforms, and with a range of technologies. However, the usefulness of existing and future datasets in answering the Mars methane question, relies on the resolutions of any detections and the overall sensitivity of the equipment.

For conventional telescopic spectroscopy, resolution is fundamentally influenced by aperture size – the bigger the telescope, the more light captured – and therefore mass. Any payload lifted off the surface of the Earth accrues a cost to overcome gravity, especially space missions which must leave the Earth's gravity well. This, balanced with competing science objectives, means that off-Earth instruments are by comparison much smaller than ground-based ones. The FLITECAM telescope aperture is 2.7 m and MRO's CRISM just 100 mm, whereas the biggest telescope on Earth, the E-ELT will be 40 m across. Other Martian in-situ technologies like MSL's TLS don't use a telescope, but clearly involve large mission costs to deliver the instrument to the target site.

Though obviating some of the constraining factors such as telluric contamination and favourable operating environments, off-Earth instruments cannot compete with on-ground facilities in terms of aperture size, accessibility, and potential for upgrading. The trade-offs for these benefits include mechanical stresses and interference, atmospheric conditions, and simple distance to target.

### 4.8.1   Spatial and Spectral Resolution

Both spatial and spectral resolution depend on a number of common parameters including slit width, optics, CCD specifications, and signal-to-noise-ratio. Modern CCDs can be used to paramatize spatial resolution along one dimension of the array, and spectral resolution along the other, making data processing a constraining factor for ultimate utility of the observation.

Close to the diffraction limit of the instrument optics, spatial resolution is a function of the number of pixels in the CCD, and their size. The more pixels the source object covers, the greater the spatial resolution. CCDs with 4096 x 4096 pixels, each of 0.004" such as those proposed for the TMT and (presumably) the E-ELT instruments will have extremely high spatial resolutions.

Spectral resolution is also a function of the number and size of pixels. The number of pixels of a certain width required to produce a Full Width Half Maximum (FWHM) spectral line, is a function of the slit width and observed wavelength, so the smaller the pixels, the greater the spectral resolution. Again, the 4096 x 4096 CCDs with very small pixels can discern the most spectral lines.

### 4.8.2    Radiometric Resolution

Radiometric resolution in astronomy is commonly based on 16-bit data rates, but overall instrument sensitivity is influenced by the Full Well Capacity of the pixels, the Data Number of the digitization process, and, especially for infra red light, the various heat noise components of the system. This makes comparison of instruments very difficult, suffice it to say, that each instrument is designed and operated for a specific range of wavelengths and for a specific science outcome. Infra red telescopes may therefore be *capable* of detecting a range of wavelengths encompassing one of interest, but may not provide valuable results for that wavelength.

A deeper investigation of radiometric resolution was deemed to be beyond the scope of this report, but in general, within the constraints of project/mission budgets, and science outcomes, radiometric resolution improves with technological advancement.

### 4.8.3    Temporal Resolution

Temporal resolution simple refers to the 'return to site' frequency of any observation. In the case of Martian atmospheric methane, this is a parameter of considerable importance because of the disputed transience of recent detections, the doubt surrounding the accuracy of the detections, and the lack of comparable, replicated results.

Ignoring other factors, and assuming a telescope has non-sidereal tracking capability, repeat observations of the same site from terrestrial instruments are constrained by the respective rotations and orbits of Earth and Mars. Optimal apparitions of Mars occur at oppositions and and at high eleveations, meaning latitude is a factor also. The oppositions occuring in 2014, 2016 and 2018 occur at increasingly southern latitudes ($5^0$, $21^0$ and $26^0$ South, respectively), the 2020 and 2022 oppositions ocurring at $5^0$ and $25^0$ North, respectively (SEDSweb, 2012).

This increasingly favours southern hemisphere, fixed instruments (the GMT and E-ELT are about $25^0$ South) for the near future, but then returns to a more favourable sky altitude for 'northern instruments' such as the TMT and IRTF facility, located on Hawaii at around $21^0$ North. Mars stays north of the equator for the subsequent seven years (SEDSweb, 2012).

# 5.    CONCLUSIONS AND RECOMMENDATIONS

Including proposed instruments, this investigation described over forty which can potentially detect at 3.3 μm – the strongest methane absorption line. Their design and specifications vary widely, in line with intended science outcomes, operational constraints, and degree of sophistication. Because of this, it was difficult to find common ground with which to compare all the spectrometers described, especially in terms of the four basic types of resolution.

Conflicting opinion was received from working astronomers about the limiting factor which determines radiometric resolution, the most common advice being that SNR was the biggest factor, and this was affected by many instrument parameters including wavelength, electronics, software, pixel type, well capacity, data number, and bit rate. The latter is commonly 16 bit in astronomical applications. Otherwise, a single parameter was not found, that could be used to compare instruments radiometrically.

Spatial and particularly spectral, resolution are most commonly considered when discussing spectroscopy, the ability to discern a target absorption line on a spectrograph being diagnostic of a species' presence. Until the development of contemporary and proposed, advanced NIR instruments, spectral and spatial resolution was complementary – space craft had better spatial resolution, and ground-based telescopes had better spectral resolution (Bailey, 2009).

Temporal resolution with respect to the Martian methane question is crucial to understanding the problem. Instruments which can temporally map the concentration of unequivocal methane concentrations will be necessary to understanding the larger atmospheric chemistry picture on Mars.

## 5.1    Capabilities and Limitations of the Three ELTs

Each of the three ELTs will have at least one NIR spectrometer with extremely high spectral resolution of over 100,000 – nearly twice that of any existing instrument. Using state-of-the-art adaptive optics systems, software, manufacturing techniques, and radiometric optimisation, they will operate near theoretical diffraction limits and be able to also achieve unprecedented spatial resolutions of around a few milliarcseconds – better than Mars-orbiting probes.

Though capable of studying solar system bodies, the three ELTs are nonetheless constrained by the exigencies of their terrestrial location. Temporal characterisation of Mars' atmosphere is tightly constrained by this fact, but still may offer better solutions than off-Earth spacecraft.

5.2     Capabilities and Limitations of Other Instruments

The range of capabilities of NIR instruments at other terrestrial, and off-Earth, locations is large and powerful and offers great opportunities to further the study of Martian methane. NIRC2, CRIRES and iSHELL will doubtless contribute significant spatial/spectral results in the near future, and FLITECAM shows particular promise with relatively cost-effective, high sensitivity observations from above the Earth's atmosphere.

Notwithstanding highly advanced equipment being readied, or currently enroute to Mars, existing spacecraft have valuable contributions to make. Mars Express and MRO instruments will provide time-dependent detections which will further the cause greatly, and the Curiosity Rover will hopefully provide very precise on-Mars data points. Space/Mars-borne missions suffer the usual constraints of any launched mission with a payload – efficacy versus cost. However, the peculiar advantages of their location complements Earth-based facilities.

5.3     Recommendations

5.3.1     that a more thorough, and exhaustive investigation be carried out into the specific capabilities of *all* existing NIR instruments with respect to their detection of methane on Mars,

5.3.2     that datasets from previous missions and facilities be mined for intrinsic or corroborating detections of Martian methane, at all of it's vibrational frequencies,

5.3.3     that the Martian atmosphere science cases for the three ELTs be given strong support, and that

5.3.4     that collaboration network between telescope project, and Mars mission teams, be strongly encouraged.

# 6.    APPENDICES

## 6.1    Literature Review

A search of the SAO/NASA (Smithsonian Astrophysical Observatory/National Aeronautics & Space Administration) abstract service revealed that since a 2003 detection of high levels of Martian atmospheric methane by a team from NASA's Goddard Space Flight Centre, there have been hundreds of peer-reviewed journal papers written on the subject. Various scientists derived clear conclusions regarding a cause, but in a recent summary of the last decade, Zahnle (2011) finds that better, more accurate science can be, and should be done, to resolve this question, both from ground-based facilities and from missions to Mars.

Although solar system atmospheres have been subject to ground- and space-based infrared spectroscopy for some time (Kostiak et al, 1999), Oliva and Origlia (2008) stated that high resolution, near infrared spectroscopy was a new area of interest and that contemporary instruments had surprisingly poor spectral resolution. They recommended a newly designed instrument for the European Extremely Large Telescope (E-ELT) which could also enhance 8-10m telescopes. This highlights the scientific novelty of this part of the spectrum, and validates the premise of the current work.

NASA's Infrared Processing and Analysis Centre (IPAC) provides an excellent introduction to infrared science (IPAC, 2012). Incident infrared radiation is absorbed by $H_2O$ and $CO_2$ in the Earth's atmosphere, though some wavelength windows exist. The atmosphere itself emits infrared radiation, compounding the problem for ground-based telescopes. The lower part of astronomers' $3.0 - 4.0$ μm 'L' Band has fairly good seeing conditions, at high altitudes (IPAC, 2012a). Telluric water vapour and methane, can be addressed "… by ratioing the spectrum of the science target by that of a calibration star…" (Gemini, 2012).

Some of methane's nine vibrational states have the same frequency, resulting in four spectral bands at about 7.3 μm, 6.3 μm, 3.3 μm and 3.2 μm (ULiv, 2012). However, Formisano et al (2004) used the Planetary Fourier Spectrometer (PFS): at around 7.7 μm, 3.3 μm, 2.3 μm and 1.7 μm. The resonant $v_3$ band at around 3.3 μm has the strongest signal to noise ratio, is most easily distinguished from other effects, and is the focus of much of the literature on Martian methane.

Telluric methane is subtracted from spectrographs using the HITRAN (high-resolution transmission molecular absorption) database which quantum characterises them (HITRAN, 2012) and FLITECAM (First Light Infrared Test Experiment CAMera) on SOFIA (Stratospheric Observatory for Infrared Astronomy) uses a version called ATRAN (FLITECAM, 2012).

Fourier Transform Infrared Spectroscopy (FTIS) allows multi-wavelength sampling, with units of $cm^{-1}$ (Griffiths & De Hasseth, 2007) and equipment is super-cooled to improve SNR. Photonic detectors being developed at Swinburne University will improve current sensitivity (Uddin, 2012).

Assisted by Adaptive Optics (AO), the theoretical, 'diffraction limit' of spatial resolution can be approached by terrestrial telescopes. Many texts and websites explain these concepts e.g. Cornell (2012), Freedman et al (2011).

Depending on orbits, Mars varies from being 56 to 400 million Km from Earth (GSFC, 2012), so spatial resolution for a 35 m ground-based telescope ranges from 115 m to 800 Km. Mars' diameter is 6760 Km, presenting an angular size 3.5" – 25" . In 2018 Mars will reach 24.3", in 2020, 22.6", in 2022, 17.2". (SEDSweb, 2012).

IUPAC (1997) defines spectral resolution (R) as the measure of a detector's ability to distinguish between two wavelengths in a spectrum; in astronomy it's expressed as a function of the target wavelength:

$$R = \frac{\lambda}{\Delta\lambda}$$

Temporal resolution refers to the rate of return to a target site. Mars Orbiters, while repeating their orbit about every one hundred minutes, do not return over the same site for many months. Compare this to a fixed telescope which can observe the same part of the Martian disk many nights consecutively. Martian rovers can only sample the immediate environment over its life time.

Pixels in Charge Couple Devices (CCDs) collect electrons up to their Full Well Capacity (FWC) - as many as 500,000 each. Radiometric resolution relates to the FWC of the CCD and the digitisation process, giving a Digital Number: DN. The FWC/DN ratio is the CCD's gain and is related to the CCD array size (Specinst, 2012; RITweb, 2012).

This is "…a very complex field…" (Brandl, 2012), detector responsivity and atmospheric background determining radiometric resolution. Astronomical instruments use 16-bit pixel resolution which is adjusted for a particular FWC for the wavelength being detected (Ryder, 2012). This was corroborated by Seppo (2012) who regards DN as the basic radiometric resolution, as opposed to Smith (2012) who uses FWC and SNR. Internal and external noise are the greater contributors to limiting radiometric resolution, and this is ameliorated by taking multiple, short detections, as well as 'dark frame' (closed aperture) detections, and averaging the result. Environmental thermal background 'noise' is addressed by locating telescopes at high, dry altitudes and using appropriate structural materials and coatings (Ryder, 2012).

Quoting a leading Mars $CH_4$ investigator: "…the history of the search for methane on Mars is long and storied" (Mumma, 2004). The infant technology of spectroscopy was applied to Mars from the mid 19[th] century (Campbell, 1894). Much, later, Kuiper et al (1947) thought that a newly designed spectrograph could help solve the nature of "…the green spots on Mars…", and Sinton Bands were at one time controversially caused by $CH_4$ (Sinton, 1957).

Contemporary detection of trace amounts of $CH_4$ in Mars' atmosphere by Mariner 9 (Maguire, 1977; Hartmann & Raper, 1974) were corroborated telescopically in 1997 (Krasnopolsky, 1997; Krasnopolsky et al, 1997). Subsequently, Summers et al (2002) suggested volcanic outgassing, Wong et al (2003) favoured biogenesis, and Caldwell et al (2003) showed how the Hubble Space Telescope (HST) could make useful observations. But it was Mumma et al (2003, 2004) who detected significant amounts with the CSHELL and Phoenix instruments on telescopes at Maunakea.

Mars Express confirmed this detection with its Planetary Fourier Spectrometer (Formisano et al, 2004; Chicarro, 2004), and Krasnopolsky et al (2004) made a similar confirmation using the Fourier Transform Spectrometer at the Canada-France-Hawaii Telescope, suggesting a biological cause. Steigerwald (2009) explained that thousands of tonnes of methane in the Martian atmosphere points to a recent process, either geological or biological.

Discussions have ranged from production of methane by methanogens in a laboratory (Kral et al, 2003), to cometary impact, hydrothermal processes, or extant micro-organisms (Atreya & Wong, 2004). Lyons et al (2005) promoted subsurface magma interactions, Krasnopolsky (2006) was unequivocal about flaws in the abiotic case, though a year later (Krasnopolsky, 2007) could not detect any methane using the same instrument as Mumma. Martin-Torres and Mlynczak (2005) suggested solar heating was a complicating factor. Mumma et al (2009) published data showing tens of thousands of tonnes of methane, Encrenaz (2008) casting doubt over the detections suggesting that "…further dedicated observations are … needed to firm up the detection …". Spacecraft detectors have been designed, promising they could successfully distinguish between biogenic and geologic methane (Wilson et al, 2007).

It was not until three years ago that high dispersion infrared spectrometers with spatial resolutions better than 500Km, were aimed at Mars (Mumma et al, 2009).  Also, because of Mars' thin atmosphere, spectral line broadening due to pressure is minimal, so high spectral resolution is required to separate atmospheric species (Encrenaz, 2008).

A Giant Magellan Telescope (GMT) science case (GMT, 2006) included studying solar system objects with 'more sensitive spectroscopy'. By 2008, the GMTNIRS (Giant Magellan Telescope Near Infra Red Spectrometer) was proposed, looking in the 3 – 5 μm range with spectral resolution up to 150,000 and spatial resolution 30 milliarcseconds (mas) (GMT, 2008). The GMT will have capacity to observe Mars in the 1 – 5 μm range with high spectral resolution and small field of view with respect to the disk of Mars (Tinney, 2012; Da Costa, 2012)

A Thirty Meter Telescope (TMT) science case (TMT, 2007) suggested their Infra Red Imager and Spectrometer (IRIS) could examine the chemical makeup of Jovian moon atmospheres so as to study the evolution of conditions suitable for life on Earth, a science driver for Martian exploration. Similar to the GMTNIRS,  the TMT will also have the Near-InfraRed Echelle Spectrometer (NIRES) (TMTweb, 2010). Although time on the telescope is 'over-subscribed', periodic, short observations of 15 minutes was a possibility (Ádámkovics, 2012).

The European Extremely Large Telescope (E-ELT) science document (Hook, 2007) explicitly mentions solar system astronomy, espousing the virtues of a telescope with spatial resolution, similar to Earth's weather satellites, proposing great benefits from temporal observations of the planets, complementing the work of spacecraft missions. It is subsequently argued that the E-ELT would make important contributions to near-by planetary astronomy, by virtue of its spatial, spectral and temporal resolution (E-ELT, 2009). The Mid-infrared E-ELT Imager and Spectrograph (METIS) will have spectral resolution up to 100,000 in the $3 - 5.3$ μm band.

Appendix 2 lists relevant existing and proposed NIR spectrometers on telescopes, facilities and missions, along with their website URLs. At the time of submission, a common set of specifications was not found for every instrument, and it should be noted that the spectrometers were/are at quite varied stages of design, construction, testing and implementation.

## 6.2       Instruments Capable of Detecting $CH_4$ at 3.3 μm

### 6.2.1    NIR Spectrometers on Earth

#### 6.2.1.1        W. M. Keck Observatory

At 4,123 m altitude on Mount Maunakea, Hawaii, Keck is operated by a consortium of the University of California, the California Institute of Technology and NASA (Keckweb, 2012). It has two 10 m diameter telescopes.

##### 6.2.1.1.1        Near Infra Red Camera (NIRC)

NIRC had a 256 x 256 pixel detector for the $0.8 - 5.5$ μm range. It had a low spectral resolution of R = 164, a field size of 5.27", and pixel size of 0.0206". It was mainly used to study galaxy formation from the Keck I telescope, but at wavelengths over 3 μm was severely limited by instrument electronics. NIRC is no longer in operation (Keckweb, 2012).

6.2.1.1.2        Near Infra Red Camera 2 (NIRC2)

NIRC2 saw first light in 2001, and has a 1024 x 1024 pixel detector for the 0.9 – 5.3 μm range. It operates at a temperature of 35 K and a specific filter to pick up 3.3 μm was installed in September 2011. It can be used to search for extra-solar planets, and also to study solar system bodies. Used in conjunction with adaptive optics, NIRC2 has the highest resolution of any ground-based instrument (Keckweb, 2012).

6.2.1.1.3        Near Infrared Spectrometer (NIRSPEC)

NIRSPEC is used on the Keck II telescope to study galaxy formation as well as the solar system. It has a 1024 x 1024 pixel detector operating in the range 0.95 – 5.5 μm, and a spectral resolution of up to 25,000. Its KL filter covers the 2.134 - 4.228 μm band (Keckweb, 2012).

6.2.1.2  Very Large Telescope Array

The Very Large Telescope Array (VLTA) is the European Southern Observatory's "flagship facility" located at 2,635 m altitude on Cerro Paranal in Chile. It consists of four 8.2 m diameter fixed scopes, and four 1.8 m moveable scopes. It is run by a collaboration of some 15 European governments, headquartered in Germany (ESOweb, 2010).

6.2.1.2.1        CRyogenic high-resolution InfraRed Echelle Spectrograph (CRIRES)

CRIRES operates from 0.92 – 5.2 μm with a spectral resolution up to 100,000. It has a 1024 x 1024 array of 0.086" pixels. It saw first light in 2006 and has a stated science goal which includes solar system objects (VLTweb, 2012).

6.2.1.2.2        Infrared Spectrometer And Array Camera (ISAAC)

ISAAC can be operated in Long Wave (2.5 – 5.0 μm) mode for a 1.25' x 1.25' field of view, and a spectral resolution of up to 10,000. It also has a 1024 x 1024 pixel array capturing 0.071"/pixel. It's science goals include faint outer solar system bodies (VLTweb, 2012).

6.2.1.3        Gemini Observatory

The Gemini Observatory has an 8.1m diameter telescope at each of the two best viewing sites in the world: Maunakea, Hawaii and Cerro Pachon, Chile. They're owned and run by an association of the United States of America, the United Kingdom, Canada, Chile, Australia, Brazil and Argentine, as well as the United States National Science Foundation (Geminiweb, 2012).

6.2.1.3.1        Gemini Near Infra Red Spectroscope (GNIRS)

GNIRS saw first light at Cero Pachon in 2004, but after damage repairs in 2007 it was relocated to Gemini North on Mt Maunakea. It has a 1024 x 1024 pixel array with spectral response 0.9 – 5.5 μm. The broadband L filter looks at 2.8 – 4.2 μm, and a narrow Band PAH (Polycyclic Aromatic Hydrocarbon) filter centres on 3.295 μm (GNIRSweb, 2012).

6.2.1.3.2        Phoenix

Though decommissioned in 2011, the Phoenix instrument's L3100 filter could just capture 3.3 μm. In 2012, it was upgraded to a 1024 x 1024 pixel array, and will achieve spectral resolution up to 70,000 over the range 1.0 – 5.0 μm (Phoenixweb, 2012). It will be used on several telescopes at the Kitt Peak National Observatory (KPNO) in Arizona, USA.

6.2.1.4        Infra Red Telescope Facility

The Infra Red Telescope Facility (IRTF) is a 3.0 m telescope located on Mt Maunakea, Hawaii, and operated by the University of Hawaii for NASA (IRTFweb, 2012).

6.2.1.4.1        Spectrograph and Imager (SPeX)

SPeX contains a 1024 x 1024 CCD array which can provide spectral resolution up to 2,000 in the range 2.3 – 5.5 μm with a field of view 30" x30" and 0.12" per pixel. It is designed for multiple wavelength detection of planetary features. It operates at around 35K and is funded by NASA and the NSF (SPeX, 2012).

### 6.2.1.4.2    National Science Foundation Camera 2 (NSFCAM2)

NSFCAM2 is an upgrade of NSFCAM. It has a 2048 x 2048 pixel array operating in the range 1.0 – 5.5 μm. Each pixel is 0.04 arcsec with a field of view of 80" x 80". NSFCAM2 has marginally better spatial resolution than SPeX, and is recommended over SPeX for L Band detections (NSFCAM2web, 2012).

### 6.2.1.4.3    Cryogenic Echelle Spectrograph (CSHELL)

CSHELL uses a 256 x 256 pixel array to achieve spectral resolutions over 40,000 in the 1.08 – 5.5 μm range. Each pixel covers 0.2". The imager covers 30 x 30 arcsec. CSHELL was used to map ozone and water in the Martian atmosphere in 2002 (CSHELL Mars, 2012).

### 6.2.1.4.4    Immersion Grating Echelle Spectrograph (iSHELL)

When completed, iSHELL will have the highest spectral resolution, 67,000, in the northern hemisphere for the range 1.15 – 5.4 μm. It will replace CSHELL and will use a 2048 x 2048 pixel array optimized in part for L band planetary science. It will use new technology which makes it smaller and lighter. Each pixel will see 0.06 arcsec with field of view 30" x 30". The spectrograph will be held at 38 K. iSHELL is expected to be able to have sufficient spectral resolution at the longer wavelengths in its range, to detect atmospheric gases on Jupiter and Saturn, through considerable telluric contamination (iSHELLweb, 2012).

### 6.2.1.4.5    Broadband Array Spectrograph System (BASS)

The Aerospace Corporation's BASS is a portable IR spectrograph covering 2.9 – 13.5 μm at low spectral resolution up to 120. In 1998, BASS was used on the 3 m IRTF telescope to take spectral images of an orange dwarf star in Hydra (Calvet et al, 2004). Recently, it was used on a 3.6 m telescope on Hawaii, to take broad spectrum thermal images 200 times per second, of the Lunar Crater Observation and Sensing Satellite (LCROSS) impact on the Moon (Heldmann et al, 2011). BASS has also been used aboard SOFIA and FISTA.

6.2.1.4.6        Midwave Infrared Imaging Spectrograph (MIRIS)

MIRIS was a successful test apparatus for semiconductor prism etching. It can be used on airborne platforms to observe meteor showers and infra red stars in the 3.0 – 5.5 μm range. Two different cameras can be used, resulting in slightly different spatial and spectral resolutions around 1.0 mrad, and 0.02 μm respectively. It has masks to allow long slit observations, and has been used on the Mt. Lemmon 60-inch telescope and the Mt. Wilson 60-inch telescopes (MIRISweb, 2012).

6.2.1.5        Giant Magellan Telescope (GMT)

Targeted for completion in 2019, groundworks for construction of the GMT began on March 23rd, 2012 (GMTblastweb, 2012). The University of Arizona is building the primary mirrors – seven of them, each 8.4 m diameter. The telescope will have a total light collection area of 368 $m^2$ and detect from 0.32 – 25 μm. Adaptive optics using a laser guide-star will take spacial resolution nearly to the theoretical limit, and spectral resolution to very high levels, and wide field multiplexing will enable large field-of-view studies (GMT, 2008). Although the stated science goals address important astronomical questions like star, galaxy and black hole formation, cosmology and dark energy, the telescope will be capable of studying solar system bodies as well (GMT, 2006).

6.2.1.5.1        Giant Magellan Telescope Near InfraRed Spectrometer  (GMTNIRS)

GMTNIRS will use immersion, echelle dispersers, detect in the 1 – 5 μm range onto a 2048 x 2048 pixel CCD, and produce a spectral resolution up to 120,000 over a 2 arcsecond field of view. The short wavelength channel will have a spectral resolution around 100,000, and the 3 – 5 μm channel will be diffraction limited at 3.5 μm. This will enable precise detections and determination of three $CH_4$ absorption lines: 1.7 μm, 2.3 μm, and 3.3 μm (GMT, 2008).

6.2.1.5.2       Mid-IR Imaging Spectrograph (MIISE)

MIISE will have two channels, one detecting in the $3 - 5$ μm range, and with adaptive optics will have a broad range of uses beyond its main purpose to study star and planet formation. It will achieve spectral resolutions around 2000 over a 2 arcminute square, area of sky, and be diffraction limited at the lower wavelength end of its range.

6.2.1.6       Thirty Meter Telescope (TMT)

Scheduled for completion around 2018, the TMT will use a single primary mirror constructed of 492 segments. With over 700 m$^2$ of light collecting area, adaptive optics based on successfully trialed Gemini technology, and instrumentation improving on Keck detectors, this telescope will be a powerful adjunct to the JWST. Achieving spatial resolutions of just a few milliarcseconds per pixel on a large 4096 x 4096 array, and spectral resolutions up to 100,000, the TMT, as well as addressing important 'big' questions in astronomy, will have the capacity to study solar system atmospheres in high detail.

6.2.1.6.1       InfraRed Imaging Spectrograph (IRIS)

IRIS will be an integral field spectrometer in the $0.85 - 2.5$ μm range, with a 17" x 17" field of view. It will have spatial resolution ten times greater than the HST, and a major science driver to study the chemical makeup of Kuiper Belt objects. Like all high-end spectrometers, IRIS will use an atmospheric dispersion corrector which inputs real-time optical and physical parameters about the intervening atmosphere, so as to correct for telluric influences. It will use optical fibre technology for multiplexed detections (IRISweb, 2012). This instrument will be capable of detecting $CH_4$ lines at 1.7 μm and 2.3 μm.

6.2.1.6.2       Near-InfraRed AO-fed Echelle Spectrometer (NIRES)

A convential echelle spectrometer, NIRES will have two channels, each achieving between 20,000 and 100,000 spectral resolution. The short band ($1.0 - 2.5$ μm), and the long band ($2.9 - 5.0$ μm) will have spatial resolution of 4 mas. NIRES will detect behind the same adaptive optics unit as IRIS (TMTEELTweb, 2012).

### 6.2.1.6.3    Wide Field AO Imager (WIRC)

If WIRC is built, it may have a field of view of 30" x 30" and low spectral resolution around 100. Designed for the range 0.6 – 5.0 µm, it would be used in conjunction with adaptive optics for a number of purposes (TMTEELTweb, 2012).

### 6.2.1.6.4    Planet Formation Instrument (PFI)

TMT's planet formation instrument is specifically designed to characterize the atmospheres of exoplanets. It will operate in the 1 – 4 µm range with low spectral resolution around 100 for a full field of view which will be one or two arcseconds radius. It will need an extremely advanced adaptive optice system (TMTEELTweb, 2012).

### 6.2.1.7    European Extremely Large Telescope (E-ELT)

Timelined to see first light "early next decade", the E-ELT will have a 39.3 m diameter primary mirror made up of 798 hexagonal segments, each 5 cm thick. This will give it 978 m$^2$ of light collecting area, the biggest in the world. Sharing site selection investigations with the TMT Corporation, Cerro Armazones, near Paranal, Chile was chosen in 2010. Over 6000 actuators will 'correct' the shape of the adaptive optics mirrors at millisecond timescales. Funding approval was recently granted to begin construction of the road to the site, and also to begin construction of one of the mirrors. Final budgetary approval for the whole project is expected in mid 2012 (ESONewsweb, 2012).

### 6.2.1.7.1    Mid-infrared Imager and Spectrograph (METIS)

METIS will be a very high spectral resolution spectrometer operating in the range 3 – 5.3 µm. With a field of view of 0.4" x 1.5", and spectral resolution up to 100,000, METIS will use adaptive optics and laser guide stars to calibrate against the atmosphere (METISweb, 2012). Stuik et al (2010) are currently designing the specifics, warning that atmospheric subtraction is a major issue.

## 6.2.2    NIR Spectrometers in Flight

### 6.2.2.1        Statospheric Observatory for Infrared Astronomy

The Stratospheric Observatory for Infrared Astronomy (SOFIA) is a modified Boeing 747-SP operated by NASA and the German Space Agency. It flies up to 45,000 ft in order to observe from above Earth's atmospheric water vapour. It has a telescope with a 2.5 m aperture (SOFIAweb, 2012).

#### 6.2.2.1.1        First Light Infrared Test Experiment CAMera (FLITECAM)

FLITECAM on SOFIA uses a 2048 x 2048 pixel array capturing in the 3.0 – 5.5 μm range. It uses a grism system and has a specific filter for PAHs at 3.3 μm. It uses an ATRAN code to subtract telluric elements (FLITECAMweb, 2012), and in 2010, it was proposed that FLITECAM could be used to characterise exoplanet atmospheres, including methane content (Angerhausen et al, 2010).

### 6.2.2.2        Flying Infrared Signatures Technology Aircraft

The United States Airforce's (USAF) Flying Infrared Signatures Technology Aircraft (FISTA) was developed in the 1970s to monitor thermal radiation in the atmosphere, from nuclear weapons testing. It has several instruments devoted to targets of interest to the US Department of Defence such as aircraft and ground installations.

#### 6.2.2.2.1        Synthetic Aperture Infra Red Spectrograph (SAIRS)

SAIRS has eight selectable filters in the range 1.3 – 5.0 μm and uses a 12 bit integrated CCD.

6.2.3    NIR Spectrometers Orbiting Earth

6.2.3.1        Hubble Space Telescope

The Hubble Space Telescope (HST) has been in orbit about 560 Km above Earth since 1990. It has a number of optical, ultra violet and infra red cameras, one of which was installed in 1997.

6.2.3.1.1        Near Infrared Camera and Multi-Object Spectrometer  (NICMOS)

NICMOS uses slitless, multi-object grism spectroscopy over the range 0.8 – 2.5 μm. It has a spectral resolution of 200 per pixel for its three cameras. NICMOS could detect the 2.3 μm and 1.7 μm methane vibrations with its 256 x 256 pixel CCD.

6.2.3.2        Spitzer Space Telescope

The Spitzer Space Telescope has been in orbit around Earth since 2003. It has an 85 cm diameter telescope cooled to under 5.5 K. It has infra red spectroscopy capabilities in the 5 – 40 μm range (Spitzerweb, 2012).

6.2.3.2.1        InfraRed Array Camera (IRAC)

Of IRAC's four cameras, one records at 3.6 μm onto a 256 x 256 array of Indium/Antimony CCD pixels. Recently, methane in the atmosphere of an extrasolar planet was tentatively detected using IRAC (Beaulieu et al, 2011).

6.2.3.3        Wide Field Infrared Survey Explorer (WISE)

In a 95 minute, polar orbit, WISE has a 40 cm diameter telescope held at a temperature of 12 - 32 K, with four cameras each with 1024 x 1024 pixels. Each pixel can see 2.75" One of the cameras receives detections at 3.4 μm, every 1.1 seconds, at a spatial resolution of 6". WISE can detect asteroids as small as 3 Km diameter and provides a source catalogue for the JWST (WISEweb, 2012).

6.2.3.4        Akari Space Observatory

Akari (Japanese for 'light') was a space observatory dedicated to an all-sky infra red survey with better resolution than IRAC. It had two instruments covering 1.7 – 180 μm, and was decommissioned in 2011 (Akariweb, 2012).

6.2.3.4.1     <u>Infra Red Camera/Near Infra Red (IRC/NIR)</u>

Akari's NIR camera was composed of three individual units. The IRC/NIR had a prism/grism disperser, and camera detected in the range 1.7 – 5.5 μm. It used a 512 x 412 pixel CCD, with spatial resolution 1.46" and field of view of 9.5' x 10' (arcminutes symbol: '). A very recent paper by Okamura et al (2012), used Akari data to look at 3.0 μm absorptions of asteroid 21 Lutetia.

6.2.3.5        Infrared Space Observatory

The European Space Agency's Infrared Space Observatory (ISO) was a dedicated infra red telescope orbiting Earth in a highly elliptical orbit. It was launched in 1995, and de-orbited in 1998. Maintained at a temperature of 2 K, it operated at 2.5 – 240 μm, and amongst other things, discovered water (ISOweb, 2012), and hydrocarbons (Fouchet et al, 2004) around solar system planets.

6.2.3.5.1     <u>Infrared Space Observatory Camera (ISOCAM)</u>

ISOCAM had two detectors of high spectral resolution, spanning the range 2.5 – 17 μm. In 1997, ISOCAM detected much 3.3 μm methane in the Jovian atmosphere (ISO, 1997).

6.2.3.5.2     <u>Infrared Space Observatory Photo-polarimeter (ISOPHOT)</u>

ISOPHOT had a low resolution photo-polarimeter in the range 2.5 – 12 μm and was used mainly to study star, galaxy formation and infra red background (ISOPHOTweb, 2012).

6.2.3.5.3    Short Wave Spectrometer (SWS)

SWS operated in the range 2.4 – 45 μm with spectral resolution up to 30,000 (SWSweb, 2012). SWS mainly detected chemical composition of interstellar space.

6.2.3.6    Arkyd 101

Planetary Resources has proposed Arkyd 101 to spectroscopically observe asteroids with the view to mining water and other compounds and elements, for use in future inter-planetary travel. It may well operate in a wavelength range including methane, and may well be able to observe Mars (Cosmiclogweb, 2012).

6.2.4    NIR Spectrometers Orbiting the Sun

6.2.4.1    Rosetta spacecraft

The European Space Agency's Rosetta spacecraft was launched in 2004 with a large suite of instruments, to observe comet 67P/Churyumov-Gerasimenko in 2014. Its flight path took at very near Mars in 2007 when it undertook some science observations, and will again shortly after its mission target is achieved (Rosettaweb, 2012).

6.2.4.1.1    Visible and Infra Red Thermal Imaging Spectrometer (VIRTIS)

VIRTIS is cooled to 70 K and uses one of its three channels solely for very high spectral resolution, echelle spectroscopy in the 2 – 5 μm range. One of its science drivers is to study gaseous species around the target comet (VIRTISweb, 2012).

6.2.5    NIR Spectrometers Orbiting Lagrange Point 2

6.2.5.1    James Webb Space Telescope

The James Webb Space Telescope (JWST) is under construction, and if launched, will settle into orbit around the Earth/Sun Lagrangian Point 2. One of the four major JWST science drivers is to study "Planetary Systems and the Origins of Life", including observations of "moving solar system targets"(JWSTweb, 2012).

### 6.2.5.1.1       Near Infrared Camera (NIRCam)

NIRCam's long wavelength, sliless, grism spectrometer will operate in the range 2.4 – 5.0 µm and be detected at a 2048 x 2048 pixel array. Each pixel will have 0.0648" resolution, for a total field 2.2' x 2.2', and spectral resolution up to 2000.

### 6.2.5.1.2       Near InfraRed Spectrograph (NIRSpec)

Operating at 30 – 40 K, NIRSpec will be the first multi-object spectrometer in space, and will have a field of view of more than 3' x 3'. It will also operate in fixed-slit, or integral-field modes, with R = 1000 at 3.0 µm.

### 6.2.5.1.3       Near-InfraRed Imager and Slitless Spectrograph (NIRISS)

NIRISS will use a 2048 x 2048 array of 18 µm pixels, and will have similar pixel resolution and field of view to NIRCam. The F356W filter wheel will allow detection from about 3.1 – 4.1 µm. The absence of a mask/slit permits maximum amount of light to enter the instrument (NIRISSweb, 2012).

### 6.2.6     NIR Spectrometers Orbiting Mars

### 6.2.6.1       Mars Express

ESA's Mars Express was launched in 2003 to "help answer fundamental questions about the geology, atmosphere, surface environment, history of water and potential for life on Mars." It has an observing period of about one hour when at pericenter of around 260 Km altitude. (MarsExpressweb, 2012).

6.2.6.1.1        Visible and Infrared Mineralogical Mapping Spectrometer (OMEGA)

This 29 Kg instrument has one channel for $1.0 - 5.2$ μm spectroscopy, with a sensitivity between $13 - 20$ nm. During the mission, the whole surface of Mars will be spectrally mapped at $1 - 4$ Km resolution, and up to 5% of the surface will be mapped at 300 m resolution. OMEGA will be targeting silicates, iron, carbonates and nitrates (OMEGAweb, 2012).

6.2.6.1.2        Planetary Fourier Spectrometer (PFS)

PFS is specifically designed to address questions about the composition of the Martian atmosphere. It will detect in a broader range of wavelengths than OMEGA ($1.2 - 45$ μm), and have better spectral resolution needed for identifying gases. PFS's short wavelength component can detect 2000 spectral points between $1.2 - 5$ μm, and if the anticipated one million spectra are collected, averaging will improve signal-noise-ratio, and produce an accurate map of the Martian atmosphere (PFSweb, 2012).

6.2.6.2        Mars Reconnaissance Orbiter (MRO)

Launched in 2005, NASA's Mars Reconnaissance Orbiter has six science instruments to conduct "global mapping, regional surveying, high-resolution trageting of specific spots" on the surface of Mars, searching for signs of water in Mars' history (MROweb, 2012).

6.2.6.2.1        Compact Reconnaissance Imaging Spectrometer for Mars CRISM

CRISM is a 100mm aperture, long-slit, scannable spectrometer, over the range $0.362 - 3.92$ μm. The long wavelength component detects $1.0 - 3.92$ μm. At 300 Km altitude, it has a swath width of about 10 Km for $60^{\circ}$ back and forth, along the ground track.

6.2.6.3        Mars Global Surveyor (MGS)

Launched in 1996 and ending in 2006, NASA's Mars Global Surveyor entered a polar orbit around Mars and sent back huge amounts of data on the topography and geology of the surface (MGSweb, 2012).

6.2.6.3.1    Thermal Emission Spectrometer (TES)

TES was an interferometer covering the range 6.25 – 50 μm, so it could have data about the longer methane wavelengths (TESweb, 2012). From 1999 to 2004, TES collected millions of methane spectra at 7.8 μm, which is enabling scientists to map temporal variations (Chizek et al, 2011).

6.2.6.4    MAVEN

MAVEN is scheduled to launch in 2013 with the specific task of characterising the Martian atmosphere, and determining how/why it has been depleted over time (MAVENweb, 2012).

6.2.6.4.1    Neutral Gas and Ion Mass Spectrometer (NGIMS)

This instrument is under development and will be one of the first to provide data to NASA in an upgraded format (Huber et al, 2012). NGIMS will help provide a "definitive statement about whether or not methane is still present in the atmosphere and characterize whatever variability it has" (Smith, 2009).

6.2.6.5    Exobiology on Mars (ExoMars)

ExoMars is a joint ESA/Roscosmos mission to send rovers and landers to Mars to look for biosignatures. It is scheduled to launch in 2016 (ExoMarsweb, 2012).

6.2.6.5.1    MicrOmega Infrared Spectrometer

MicrOmega has 20 μm x 20 μm pixels and can operate in broad spectrum mode from 0.9 – 3.5 μm. It is in prototype design and testing phase (NGIMSweb, 2012), and should be able to detect below 10 ppt (Zurek et al, 2010).

6.2.6.5.2        Raman Spectrometer

Also in the design and testing phase, the Raman Spectrometer for ExoMars is likely to be able to detect at 3.3 μm (Ramanweb, 2012).

6.2.7    NIR Spectrometers on Mars

6.2.7.1        Mars Science Laboratory (MSL)

The Mars Science Laboratory will land on Mars in August, 2012. It is tasked with meeting eight main objectives about Mars' biological, geological, planetary, and radiative properties. It carries four spectrometers including the Sample Analysis at Mars (SAM) suite (SAMweb, 2012).

6.2.7.1.1        Tunable Laser Spectrometer (TLS)

The TLS will have parts per trillion sensitivity to detect $CH_4$ at 3.27 μm range (Mahaffy et al, 2012), by using an improved version of Raman spectroscopy (TLSweb, 2012). SAM was tested and calibrated in a Mars-like environment (Mahaffy et al, 2011), and will analyse sedimentary layers in Gale Crater.

6.2.7.2        GOWON and BOLD

Gone With The Wind On Mars (GOWON) is a 'blue sky' concept whereby a Mars-orbiting probe handles networked data from a multitude of lightweight, wind-blown sensors released on Mars (Davoodi et al, 2012). It is similar in overarching concept to the Biological Oxidant and Life Detection (BOLD) proposal which would drop a number of pyramid-shaped science stations onto the surface of Mars, and send detection results to an orbiter (WSUweb, 2012).